\documentclass[conference]{IEEEtran}
\usepackage{cite}
\usepackage{amsmath,amssymb,amsfonts}
\usepackage{algorithmic}
\usepackage{graphicx}
\usepackage{textcomp}
\usepackage{xcolor}
\usepackage{float}
\usepackage{tikz}
\usepackage{eso-pic}
\usetikzlibrary{automata, positioning, arrows}
\def\BibTeX{{\rm B\kern-.05em{\sc i\kern-.025em b}\kern-.08em
    T\kern-.1667em\lower.7ex\hbox{E}\kern-.125emX}}

\AddToShipoutPictureBG*{
  \AtPageUpperLeft{
    \put(0,-40){
      \makebox[\paperwidth][c]{
        \footnotesize 
        \color{black} Accepted for presentation at the 45th AIAA/IEEE DASC, Orlando, FL, USA, 2026. \copyright 2026 IEEE. Personal use permitted.
      }
    }
  }
}

\begin{document}
\title{A Safety-Driven Architectural Framework for Fail-Operational Drone Swarms in Critical Missions}
\author{\IEEEauthorblockN{Luiz Giacomossi, Zafer Yigit, Marwan Shakarna, 
Shoaib Saleemi, Ivan Tomasic, \\ Baran \c{C}ur\"{u}kl\"{u}, and Håkan Forsberg}
\IEEEauthorblockA{\textit{Mälardalen University} \\
Västerås, Sweden \\
Email: luiz.giacomossi@mdu.se}
}

\maketitle

\begin{abstract}
The certification of Unmanned Aerial Vehicle (UAV) swarms for safety-critical operations requires verifiable design assurance. Airworthiness standards demand deterministic reliability, whereas multi-agent coordination algorithms execute non-deterministic models. This paper proposes a mixed-criticality architectural framework that applies SAE ARP4754B methods to swarm reconfiguration. First, a hardware-isolated \textit{Safety Monitor} functions as a Run-Time Assurance (RTA) gateway, decoupling the flight-critical core from the non-deterministic Swarm Manager. Second, the monitor enforces formal safety contracts based on agent Health Vectors derived systematically from a Functional Hazard Assessment (FHA). Third, the framework propagates these Health Vectors to the collective planner to trigger fail-operational task reallocation, enabling intelligent swarm behaviors without compromising flight-critical isolation. Markov reliability modeling demonstrates that the $10^{-7}$ failures per flight hour Hazardous target is theoretically achievable for our SAIL~IV scenario, provided the Safety Monitor meets $C_{monitor} > 0.9991$, consistent with DAL~B CMD/MON implementations.
\end{abstract}

\begin{IEEEkeywords}
Fail-Operational Systems, UAV Swarms, Safety-Driven Architecture, Swarm Resilience, Safety-Critical Systems
\vspace{-3mm}
\end{IEEEkeywords}

\section{Introduction}
\label{sec:introduction}
%% ========================
%% Introduction
%% ========================
Unmanned Aerial Vehicle (UAV) swarms in safety-critical missions, such as Search and Rescue (SAR), require operational authorization based on formal design assurance \cite{Chung2018, search_LARS, Search_IAI}. While coordinated multi-agent systems offer increased area coverage over single agents, their non-deterministic emergent behaviors pose a certification challenge. As the primary regulatory framework for civil UAVs, the JARUS Specific Operations Risk Assessment (SORA) imposes quantitative reliability targets. Our reference scenario, a Beyond Visual Line of Sight (BVLOS) search over a sparsely populated area, corresponds to medium-risk SAIL~IV. At this level, SORA mandates compliance with recognized airworthiness standards; we therefore adopt civil aviation severity targets, where a \emph{Hazardous} condition (e.g., an uncontrolled crash) must remain below $10^{-7}$ per flight hour \cite{JARUS_SORA}. Achieving this target requires a verifiable link between system architecture and operational safety \cite{Johnson1984}.

An architectural gap exists between single-agent avionics and multi-agent robotics. The fault-tolerant systems community prioritizes deterministic, fail-safe designs using redundant hardware and time-triggered buses to mitigate failures \cite{Zhang2021}. Conversely, swarm robotics research focuses on collective resilience. Recent works have proposed "health-aware" coordination strategies \cite{Zhao2025, Alamdar2025}. However, these approaches typically optimize for mission performance (e.g., energy efficiency) rather than design assurance. They lack the formal isolation required to prevent a non-deterministic replanning algorithm from violating safety constraints. This creates a mixed-criticality integration problem: safety-critical flight control laws must coexist with complex, unverified swarm logic, requiring a high-integrity mediator to guarantee isolation \cite{ASTM_F3269}.

This paper proposes a safety-driven architectural framework toward fail-operational drone swarms. We apply the SAE ARP4754B systems engineering methodology, the civil aviation guideline for aircraft and systems development assurance, to derive a Run-Time Assurance (RTA) mechanism aligned with ASTM F3269-17, a standard practice for safely bounding the behavior of UAV with complex functions \cite{ASTM_F3269}, that bounds swarm behavior. The contributions of this work are:
\begin{enumerate}
    \item A verifiable health-aware architecture that bridges ad-hoc swarm resilience and formal safety standards. Unlike traditional approaches using software self-reporting, we introduce a hardware-isolated Safety Monitor to enforce the accuracy of the health state, guaranteeing the swarm's reconfiguration is based on trusted data.
    \item A systematic derivation of agent health states from Functional Hazard Assessment (FHA), with traceability from component faults to swarm reconfiguration behaviors.
    \item A quantitative reliability analysis using Markov modeling to identify the Safety Monitor coverage requirements necessary to satisfy the $10^{-7}$ probability of failure target for hazardous operations.
\end{enumerate}

While the framework is illustrated via a SAR mission under SORA SAIL~IV, the architectural principles for isolating monitoring from planning apply to any safety-critical multi-agent system with mixed-criticality partitioning.

This paper presents a conceptual framework and a theoretical safety argument. The evaluation relies on systems engineering methods appropriate for this design phase: architectural traceability, Markov reliability modeling, and a ConOps scenario walkthrough. Hardware implementation, formal software verification, and flight-test validation are future work and remain outside this contribution's scope.

\section{Related Work}
\label{sec:related_work}
% ====================================================================
% SECTION 2: RELATED WORK
% ====================================================================

The design of dependable systems for regulated airspace relies on established principles of fault tolerance through redundancy \cite{Johnson1984}. Our work connects fault-tolerant avionics, resilient swarm coordination, and RTA for complex systems.

%\subsection{Single-Agent Fault Tolerance}
Architectures for fault-tolerant UAVs traditionally focus on fault containment within a single platform. These designs employ hardware replication, such as Triple-Modular Redundancy (TMR) \cite{Wu2017}, or analytical redundancy through robust control. Approaches range from controllers that maintain stability after actuator failures \cite{Nguyen2019, Mueller2014} to energy management systems optimizing component longevity \cite{Gao2023}. While these methods provide a rigorous foundation for \textit{fail-safe} agents, they inherently treat the individual UAV as the system boundary. In these architectures, if a fault exceeds the local recovery capability, the agent typically executes a solitary emergency landing. In a SAR context, this creates an unmitigated coverage gap, as the remaining agents are unaware of the safety-critical state change.

%\subsection{Swarm Resilience}
Multi-agent systems research focuses on collective resilience, ensuring mission completion despite agent loss. Strategies include centralized replanning to restore connectivity \cite{CaregnatoNeto2022,giacomossi2026marketbasedreplanningsafetycriticaluav } or distributed self-healing for communication repair \cite{Varadharajan2020}. Recently, "health-aware" strategies have emerged, where swarm behavior adapts to specific health metrics. For instance, Zhao et al. \cite{Zhao2025} present a path planner that recalculates flight parameters after actuator faults, while Alamdar and Petrović \cite{Alamdar2025} propose connectivity maintenance based on battery depletion. However, these approaches are primarily \textit{performance-driven}. Crucially, they introduce a dependency on the very software layer that may be compromised. They operate under the assumption that the degraded agent can still reliably execute complex negotiation algorithms. If a physical fault is accompanied by a software anomaly (e.g., thread deadlock), the health propagation fails. They lack a formal safety assessment to guarantee that the reconfigured behavior does not violate airspace safety constraints, rendering them difficult to certify under deterministic standards like ARP4754B.

%\subsection{Run-Time Assurance (RTA)}
To certify non-deterministic algorithms on UAVs (or other systems not typically certified under traditional aircraft guidance regulations), the avionics community has adopted Run-Time Assurance (RTA) architectures \cite{torens2024certification, ASTM_F3269}. This concept is grounded in the logical Simplex Architecture \cite{Sha2001}, which wraps complex, untrusted functions (e.g., AI planners) with a verifiable safety monitor. Bak et al. extended this to the \textit{System-Level Simplex} \cite{Bak2009}, moving the safety monitor to isolated hardware (e.g., FPGA) to protect against Operating System (OS) failures. While these architectures successfully protect single UAVs from software faults, their application to multi-agent systems remains unexplored. Existing RTA frameworks enforce local ``fail-safe'' maneuvers (e.g., hover or terminate) but lack the mechanisms to propagate hardware-verified safety states to the collective for ``fail-operational'' reconfiguration.

%\subsection{Gap Analysis}
A gap exists between these domains. System-Level Simplex \cite{Bak2009} provides rigorous hardware isolation but lacks swarm awareness. Health-aware swarm strategies \cite{Zhao2025} provide flexibility but lack determinism. Our work addresses this by proposing a hardware-enforced, \textit{compliance-driven architectural framework}. We extend the System-Level Simplex concept by treating the Swarm Coordination Module as an "Untrusted Complex Function." We use the SAE ARP4754B safety process to derive a formal interface between the agent's isolated safety monitor and the collective logic, establishing a traceable chain of resilience from component-level faults to mission-level reconfiguration.

\section{Safety-Driven Design Methodology}
\label{sec:methodology}
% ====================================================================
% SECTION III: METHODOLOGY (FINAL CALIBRATED VERSION)
% ====================================================================

\begin{figure}[ht]
    \centering
    \includegraphics[width=0.95 \columnwidth]{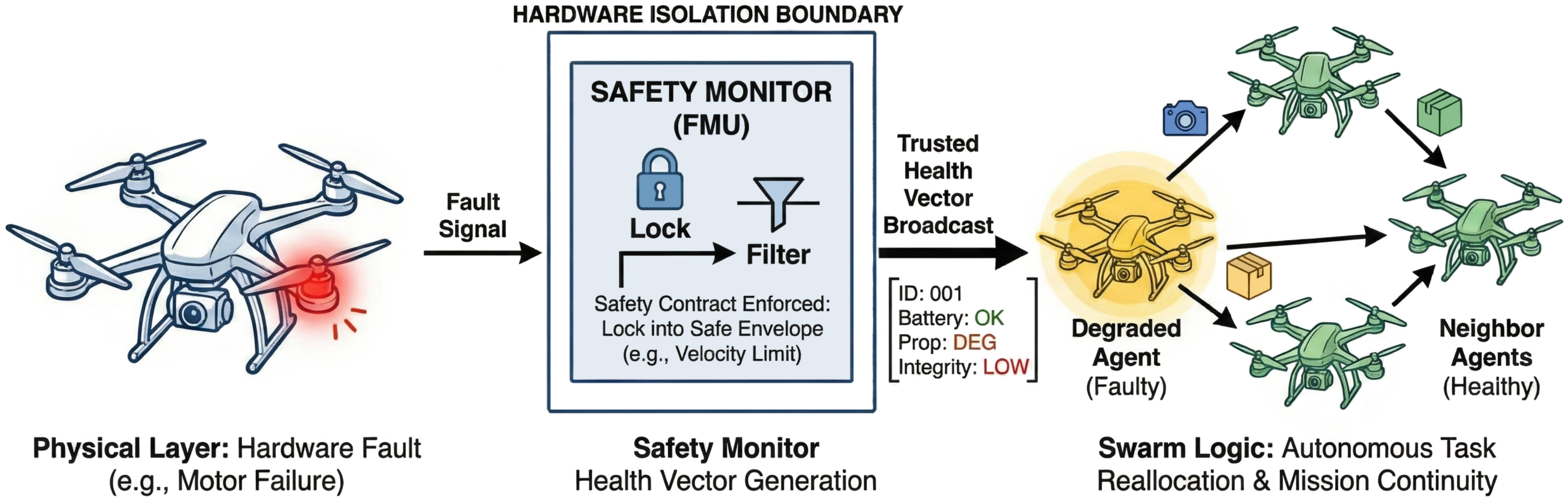} 
    \vspace{-3mm}
    \caption{Data flow of the Verifiable Health-Aware Framework. 
    The architecture acts as a \textit{Hardware Isolation Boundary} between physical and logical domains. 
    (1) Physical Layer: Component faults (e.g., a propulsion degraded motor failure) are detected by the Safety Monitor. 
    (2) Safety Monitor: The FMU uses a ``Lock and Filter'' mechanism to constrain control and filter fault data. 
    (3) Swarm Logic: The resulting ``Health Vector'' (e.g., $\langle\text{NOMINAL}, \text{DEGRADED}, \text{NOMINAL}\rangle$, Eq.~\ref{eq:health_vector}) is broadcast to the collective, triggering autonomous reallocation of mission tasks to healthy neighbors (icons show logical transfer).}
    \label{fig:concept_flow}
\end{figure}

The proposed architecture is derived from a systematic safety assessment to support certification. Our methodology adapts SAE ARP4761A~\cite{ARP4761A}, the civil aviation guideline for safety assessment of airborne systems, and ARP4754B~\cite{SAE_ARP4754B} to the constraints of mixed-criticality swarms. Unlike traditional single-aircraft processes, we explicitly analyze the propagation of faults originating in the physical agent and impacting the collective behavior, as seen in Fig.~\ref{fig:concept_flow}.

% ====================================================================
% FIGURE A: DATA FLOW 
% ====================================================================

To define the architectural components and interfaces shown in Fig.~\ref{fig:concept_flow}, the design process follows a top-down hierarchy:
1) Hazard Identification (HazID) to establish swarm-level safety bounds; 
2) Functional Hazard Assessment (FHA) to map functions to Design Assurance Levels (DAL); and 
3) Formal Interface Definition, applying Contract-Based Design principles~\cite{benveniste2018contracts} to mathematically derive the health states required to contain the identified hazards.

% --------------------------------------------------------------------
% SUBSECTION: HAZARD ANALYSIS
% --------------------------------------------------------------------
\subsection{Hazard Analysis and Risk Class}
\label{ssec:hazards}

A Preliminary Hazard Identification (HazID) established the top-level safety constraints. We align our severity classifications with the JARUS Specific Operations Risk Assessment (SORA)~\cite{JARUS_SORA}. We assume a Search and Rescue mission in a sparsely populated environment, corresponding to SAIL IV (Specific Assurance and Integrity Level 4). 

In this context, the uncontrolled crash of a single drone (H-02) is classified as a Hazardous event (Target Prob. $<10^{-7}$/hr). Although the immediate ground risk in sparse areas is low, a \textit{Loss of Control} event in a Beyond Visual Line of Sight (BVLOS) swarm allows the agent to deviate from its operational geofence, potentially infringing on civil airspace or threatening the remaining fleet (precursor to H-01). Thus, a Hazardous integrity target is required for containment. Table~\ref{tab:hazards} summarizes critical hazards and probability targets.

\begin{table}[htbp]
\caption{System-Level Hazards (Aligned with JARUS SORA SAIL IV)}
\label{tab:hazards}
\vspace{-3mm}
    \begin{center}
    \resizebox{\columnwidth}{!}{%
        \begin{tabular}{|c|p{5.5cm}|c|c|}
        \hline
        \textbf{ID} & \textbf{Hazard Description} & \textbf{Severity} & \textbf{Target Prob.} \\
        \hline
        \hline
        H-01 & Mid-air or terrain collision of multiple drones. & Catastrophic & $<10^{-9}$/hr \\
        \hline
        H-02 & Uncontrolled crash of a single drone. & Hazardous & \textbf{$<10^{-7}$/hr} \\
        \hline
        H-03 & Failure to detect the search target (Mission Failure). & Minor & $<10^{-3}$/hr \\
        \hline
        H-06 & Loss of Command and Control (C2) link. & Major & $<10^{-5}$/hr \\
        \hline
        H-07 & Unauthorized control of the swarm. & Catastrophic & $<10^{-9}$/hr \\
        \hline
        \end{tabular}
    }
    \end{center}
    \vspace{-5mm}
\end{table}

%\vspace{-4mm}

% --------------------------------------------------------------------
% SUBSECTION B: FUNCTIONAL HAZARD ANALYSIS
% --------------------------------------------------------------------
\subsection{Functional Hazard Assessment (FHA)}
\label{ssec:fha}
The FHA maps top-level hazards to specific functional failures and assigns a Design Assurance Level (DAL) to each (Table~\ref{tab:fha}). This partitions the mixed-criticality system: while top-level flight hazards require DAL~B integrity, the use of dual-channel redundancy allows the flight control modules to be developed at DAL~C (see Tab.~\ref{tab:fha}-footnote a). The Safety Monitor (FMU) is assigned to DAL~B consistent with its Hazardous failure classification, and non-deterministic swarm functions are assigned to DAL~D. The FHA includes the failure modes of the Fault Management Unit (FMU) to evaluate the integrity requirements of the Run-Time Assurance (RTA) gateway against Hazardous safety targets.

\begin{table*}[!ht]
\caption{Functional Hazard Assessment (FHA) with Calibrated Design Assurance Levels (DAL)}
\label{tab:fha}
\centering
\resizebox{\textwidth}{!}{%
    \begin{tabular}{|c|p{4.7cm}|p{6.0cm}|c|c|}
        \hline
        \textbf{Function} & \textbf{Failure Condition} & \textbf{Effect on System/Mission} & \textbf{Severity} & \textbf{Target FDAL} \\
        \hline \hline
        \textbf{F1: Safety Monitoring (FMU)} & Loss of isolation, missed fault detection, or spurious contract enforcement & Uncontained swarm faults propagate to flight core; risk of uncontrolled crash (H-02). & Hazardous & \textbf{DAL B} \\
        \hline
        \textbf{F2: Flight \& Navigation}\textsuperscript{a} & Loss of ability to determine state (position/attitude) & Uncontrolled descent; potential crash (H-02). & Hazardous & \textbf{DAL B} \\
        \hline
        \textbf{F3: Target Detection} & Failure to detect valid target (False Negative) & Reduced search effectiveness (H-03). & Minor & \textbf{DAL D} \\
        \hline
        \textbf{F4: Swarm Coord. Broadcast}\textsuperscript{b} & Transmission of corrupted trajectory data & Risk of inter-agent collision (H-01) if not validated. & Hazardous & \textbf{DAL D (B with monitor)} \\
        \hline
        \textbf{F5: Power Management} & Incorrect State of Charge (SoC) reporting & Unexpected depletion and crash (H-02). & Major & \textbf{DAL C} \\
        \hline
    \end{tabular}
    }
    \vspace{1mm}\\
    \parbox{\textwidth}{\footnotesize
    \textsuperscript{a} Although total navigation loss constitutes Hazardous condition, it requires the simultaneous failure of two independent channels. With each channel developed to $\lambda_{ch} \approx 10^{-4}$/hr, the probability of simultaneous dual-channel failure is $\lambda_{ch}^2 \approx 10^{-8}$/hr, one order of magnitude below Hazardous ($10^{-7}$/hr), satisfying the requirement at the system level. Partial navigation degradation (single-channel loss) is handled by the $\mathcal{C}_{nav}$ contract (Section~\ref{ssec:formal_interface}), which forces a controlled landing before total state loss can occur, preventing the fault from progressing to the Hazardous condition in practice.\\
    \textsuperscript{b} F4 is DAL~D because the DAL~B Safety Monitor validates all trajectory commands via the Safety Contract (Section~\ref{ssec:formal_interface}), containing the effect of corrupted swarm data. The residual probability of F4 causing H-01 is $P(\text{H-01} \mid \text{F4}) \le \lambda_{F4} \cdot (1 - C_{monitor})$. With $C_{monitor} > 0.9991$, this is below the $10^{-7}$/hr Hazardous threshold, supporting DAL~D per ASTM~F3269~\cite{ASTM_F3269}. While ASTM F3269 supports monitoring of unassured functions (DAL E), ARP 4754B~\cite{SAE_ARP4754B} specifies that DAL E functions do not require a structured development process; therefore, we do not consider this level for any subsystem.\\
    \textit{General Note: The FHA covers functions directly relevant to the proposed mixed-criticality architecture. DAL~A (Catastrophic single-point failure) is absent because Catastrophic hazards (e.g., H-01) are mitigated via architectural redundancy, reducing subsystem requirements to DAL~B/C.}
    }
    \vspace{-3mm}
\end{table*}

%\textsuperscript{a} F2 is Major rather than Hazardous through architectural mitigation argument. Although total navigation loss constitutes Hazardous condition, it requires the simultaneous failure of two independent channels. With each channel developed to DAL~C ($\lambda_{ch} \approx 10^{-4}$/hr), the probability of simultaneous dual-channel failure is $\lambda_{ch}^2 \approx 10^{-8}$/hr, one order of magnitude below Hazardous ($10^{-7}$/hr), satisfying the requirement at the system level. Partial navigation degradation (single-channel loss) is handled by the $\mathcal{C}_{nav}$ contract (Section~\ref{ssec:formal_interface}), which forces a controlled landing before total state loss can occur, preventing the fault from progressing to the Hazardous condition in practice.\\\textsuperscript{b} F4 is DAL~E because the DAL~B Safety Monitor validates all trajectory commands via the Safety Contract (Section~\ref{ssec:formal_interface}), containing the effect of corrupted swarm data. The residual probability of F4 causing H-01 is $P(\text{H-01} \mid \text{F4}) \le \lambda_{F4} \cdot (1 - C_{monitor})$. With $C_{monitor} > 0.999$, this is below the $10^{-7}$/hr Hazardous threshold, supporting DAL~E per ASTM~F3269~\cite{ASTM_F3269}.\\ \textit{General Note: The FHA covers functions directly relevant to the proposed mixed-criticality architecture.}

% --------------------------------------------------------------------
% SUBSECTION C: FORMAL DERIVATION OF SAFETY CONTRACTS
% --------------------------------------------------------------------
\subsection{Formal Derivation of Safety Contracts}
\label{ssec:formal_interface}

To decouple the DAL~B Safety Monitor and the deterministic Flight Core (DAL~C) from the non-deterministic Swarm Coordinator (DAL~D), the architecture employs Contract-Based Design (CBD)~\cite{benveniste2018contracts}. This formalism establishes a set of Safety Contracts $\mathbf{C} = \{ \mathcal{C}_{nav}, \mathcal{C}_{prop}, \mathcal{C}_{comm} \}$ that bound the permissible behavior of the swarm.

A contract $\mathcal{C}$ for a component $M$ is defined as a pair $(A, G)$, where $A$ represents the \textit{Assumptions} regarding the environment and inputs, and $G$ denotes the \textit{Guarantees} provided by the component. Component $M$ satisfies $\mathcal{C}$ if it behaves according to $G$ whenever the environment satisfies $A$.

\subsubsection*{Hierarchical Fault Abstraction and Health Vector}
To prevent the swarm algorithms processing hardware-specific fault data, the architecture implements a three-tier hierarchical abstraction:
\begin{enumerate}
    \item \textbf{Atomic Level:} Physical sensors detect localized hardware errors (e.g., motor RPM deviation, battery voltage $V_{batt}$ drop, GNSS dilution, packet loss).
    \item \textbf{Functional Level:} The FMU evaluates atomic errors against predefined safety thresholds, mapping them to a discrete functional state $\sigma \in \{\text{NOMINAL}, \allowbreak\text{DEGRADED}, \allowbreak\text{FAILED}\}$.
    \item \textbf{Aggregated Level:} The functional states of subsystem categories are aggregated into the \textit{Health Vector} $H_i$ for agent $i$:
\end{enumerate}
\begin{equation} \label{eq:health_vector}
    H_i(t) = \langle \sigma_{nav}, \sigma_{prop}, \sigma_{comm} \rangle
\end{equation}

\begin{table}[htbp]
\caption{Hierarchical Mapping of Atomic Faults to Health Vector States}
\label{tab:atomic_mapping}
\centering
\resizebox{\columnwidth}{!}{%
\begin{tabular}{|l|p{4cm}|p{1.5cm}|p{1.8cm}|}
\hline
\textbf{Atomic Signal} & \textbf{Threshold Condition} & \textbf{Functional State ($\sigma$)} & \textbf{Health Vector Component} \\ \hline \hline
RPM mismatch & $|\text{RPM}_{cmd} - \text{RPM}_{meas}| > \epsilon_{RPM}$ & DEGRADED & $\sigma_{prop}$ \\ \hline
Battery voltage & $V_{batt} \le V_{crit}$ & DEGRADED & $\sigma_{prop}$ \\ \hline
GNSS HDOP & $\text{HDOP} > \text{HDOP}_{max}$ & FAILED & $\sigma_{nav}$ \\ \hline
Heartbeat timeout & $t - t_{last\_hb} > \tau_{timeout}$ & FAILED & $\sigma_{comm}$ \\ \hline
SNR & $\text{SNR} < \text{SNR}_{min}$ & FAILED & $\sigma_{comm}$ \\ \hline
\end{tabular}%
}
\vspace{-2mm}
\end{table}

For the ConOps analysis (Section~V-D), $H_i$ is instantiated for specific agents by substituting the agent identifier, e.g., $H_D(t)$ for the faulted agent D.

\subsubsection*{Timing and Latency Bounds}
To guarantee temporal determinism for the RTA gateway, the Health Vector update is bounded by strict timing constraints. The FMU evaluates $H_i(t)$ at a fixed update frequency $f_{update}$. Upon the occurrence of an atomic fault at $t=0$, the total latency for fault detection ($\tau_{detect}$) and subsequent contract enforcement ($\tau_{enforce}$) must satisfy the maximum latency bound $\tau_{max}$:
\begin{equation} \label{eq:latency}
    \tau_{detect} + \tau_{enforce} \le \tau_{max}
\end{equation}
The bound $\tau_{max}$ is derived from the vehicle's time-to-divergence under a worst-case fault. For a hexacopter with a single-motor winding fault producing a thrust asymmetry $\Delta F$, the resulting uncompensated angular acceleration is $\ddot{\phi}_{fault} = \Delta F \cdot l_{arm} / I_{xx}$, where $l_{arm}$ is the motor arm length and $I_{xx}$ is the roll-axis moment of inertia. The time to reach the maximum recoverable attitude deviation $\Delta\phi_{max}$ (nominally $30^\circ$ for a hexacopter before loss of controllability~\cite{Mueller2014}) is:
\begin{equation}
\resizebox{0.36\columnwidth}{!}{$
    \tau_{max} = \sqrt{\frac{2 \cdot \Delta\phi_{max}}{\ddot{\phi}_{fault}}}
$}
\end{equation}

For representative platform parameters (e.g., $10$~kg mass, $l_{arm}=0.5$~m, maximum thrust asymmetry $\Delta F=60$~N, and $I_{xx} \approx 0.25$~kg$\cdot$m$^2$), this yields $\tau_{max}$ in the range of 80--150~ms, supporting the 100~ms design target used throughout this work. The specific value is a conservative platform-level requirement to be verified during hardware integration.

\subsubsection*{Contract Instantiation}
The contracts mathematically formalize the mapping of atomic faults into functional degradation. The \textit{Propulsion Contract} ($\mathcal{C}_{prop}$) restricts actuation authority when an actuator fault or a critical power degradation is detected:
\begin{equation}
\resizebox{0.91\columnwidth}{!}{$
    \mathcal{C}_{prop} : \begin{cases} 
    A: \text{FMU Powered} \land \text{Sensors Active} \\
    G: ((| \text{RPM}_{cmd} - \text{RPM}_{meas} | > \epsilon_{RPM}) \lor (V_{batt} \le V_{crit})) \implies \dots \\ 
    \quad \quad (\sigma_{prop} \leftarrow \text{DEGRADED}) \land (\|\mathbf{v}_{cmd}\| \le v_{safe})
    \end{cases}
$}
\end{equation}
where $\epsilon_{RPM}$ is the kinematic error threshold, $V_{batt}$ is the telemetry provided by the Battery Management System (BMS), $V_{crit}$ is the minimum voltage threshold to prevent brownout during high-thrust transients, and $v_{safe}$ is the resulting velocity limit imposed.

The \textit{Navigation Contract} ($\mathcal{C}_{nav}$) mandates a fail-safe transition upon sensor uncertainty:
\begin{equation}
\resizebox{0.9\columnwidth}{!}{$
    \mathcal{C}_{nav} : \begin{cases} 
    A: \text{FMU Powered} \land \text{Sensors Active} \\
    G: (\text{GNSS}_{HDOP} > \text{HDOP}_{max}) \implies \dots \\ 
    \quad \quad (\sigma_{nav} \leftarrow \text{FAILED}) \land (\text{Mode} \leftarrow \text{LAND})
    \end{cases}
$}
\end{equation}
where $\text{HDOP}$ is the Horizontal Dilution of Precision (a standard metric for GNSS positioning quality), and $\text{HDOP}_{max}$ is the threshold for safe navigation. Finally, the \textit{Communication Contract} ($\mathcal{C}_{comm}$) addresses Link Loss (H-06) by monitoring the telemetry heartbeat and Signal-to-Noise Ratio (SNR):
\begin{equation}
\resizebox{0.91\columnwidth}{!}{$
    \mathcal{C}_{comm} : \begin{cases} 
    A: \text{FMU Powered} \land \text{COM Hardware Active} \\
    G: ((t - t_{last\_hb} > \tau_{timeout}) \lor (\text{SNR} < \text{SNR}_{min})) \implies \dots \\ 
    \quad \quad (\sigma_{comm} \leftarrow \text{FAILED}) \land (\text{Mode} \leftarrow \text{RTB})
    \end{cases}
$}
\end{equation}
where $t_{last\_hb}$ is the timestamp of the last valid heartbeat from the swarm mesh, $\tau_{timeout}$ is the maximum allowable communication delay, and $\text{SNR}_{min}$ is the threshold for link integrity. If the condition is met, the FMU enforces a Return-To-Base (RTB) maneuver.

\textbf{Contract Coverage:} The three contracts jointly provide necessary coverage of the failure paths leading to H-02 at the component fault level. $\mathcal{C}_{prop}$ addresses propulsion-related faults (actuator anomalies) and power anomalies (F5) through combined RPM mismatch and $V_{batt}$ monitoring, the latter fed directly from the BMS to the FMU, as defined in Section~\ref{sec:architecture}. $\mathcal{C}_{nav}$ addresses navigation state degradation (F2) by gating on GNSS quality and transitioning to a controlled landing before total state loss. $\mathcal{C}_{comm}$ addresses link loss (H-06) by enforcing RTB before communication failure can lead to uncontrolled flight. Coverage of compound failures—simultaneous multi-subsystem faults—is addressed at the architectural level through hardware redundancy in the Flight-Critical Core and is analyzed in the FTA (Section~V-A). The contracts represent necessary but not sufficient conditions for system safety; sufficiency at the implementation level requires formal verification of the FMU logic, identified as future work.

This formulation defines the Guarantee ($G$) of each contract as a direct algebraic link between physical anomalies and the resulting discrete state $\sigma$, satisfying the hierarchical abstraction requirement while preserving traceability to the DAL B hardware monitors.

% --------------------------------------------------------------------
% SUBSECTION: SYSTEM SAFETY REQUIREMENTS (SSRS)
% --------------------------------------------------------------------
\subsection{System Safety Requirements (SSRs)}
\label{ssec:ssrs}
Based on the FHA and Safety Contracts, we derive the following SSRs that drive the architectural design:

\begin{itemize}
    \item \textbf{[SSR-FLT-01]} The Flight Critical Core shall maintain DAL C integrity for attitude estimation and control.
    \item \textbf{[SSR-FTR-01]} The system shall implement a hardware-isolated Safety Monitor of DAL B integrity to enforce the Safety Contract defined in Sec.~\ref{ssec:formal_interface}. The monitor shall achieve a diagnostic coverage $C_{monitor} > 0.9991$ for flight-critical faults, as derived in Section~V-B.
    \item \textbf{[SSR-FTR-02]} No single point of failure in the DAL~D Swarm Coordination Module (after validation by the Safety Monitor) shall cause a hazardous or catastrophic event.
    \item \textbf{[SSR-MSN-02]} The swarm logic shall autonomously re-allocate tasks when an agent reports a $\sigma \neq \text{NOMINAL}$ state.
    \item \textbf{[SSR-SEC-01]} All health vector transmissions shall be authenticated to prevent unauthorized access (H-07).
\end{itemize}

\section{Fail-Operational Architecture}
\label{sec:architecture}
% ====================================================================
% SECTION IV:  FAIL-OPERATIONAL ARCHITECTURE
% ====================================================================

To satisfy the SSRs, we propose a multi-level fail-operational architecture that addresses fault tolerance at two layers: the individual agent and the collective swarm. This hierarchy provides local fault containment while the swarm maintains mission continuity.

% --------------------------------------------------------------------
% SUBSECTION A: INDIVIDUAL AGENT ARCHITECTURE
% --------------------------------------------------------------------
\subsection{Individual Agent Architecture}
\label{ssec:agent_arch}

\begin{figure}[htbp]
    \centering
    \includegraphics[width=0.9\linewidth]{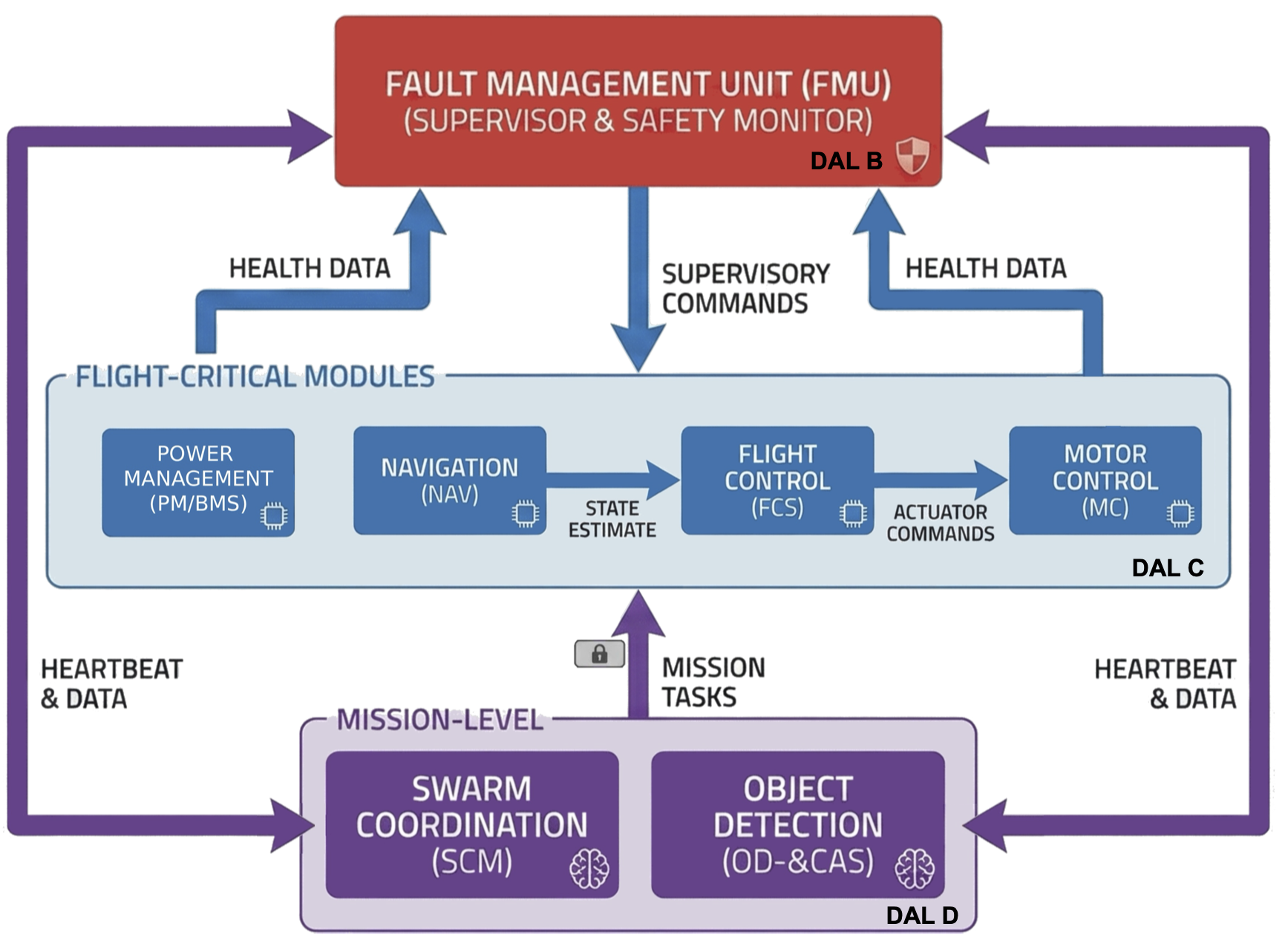}
     \vspace{-2mm}
    \caption{\textbf{Functional Control Architecture.} Logical view of safety hierarchy and DAL boundaries. The FMU is a high-integrity Run-Time Assurance (RTA) gateway (DAL B). To prevent untrusted DAL D \textit{Mission Tasks} from violating constraints, the FMU applies a ``Lock and Filter'' mechanism (denoted by the padlock symbol): \textit{Supervisory Commands} override mission inputs in the FCS if a safety contract breaches. Bidirectional links enable FMU subsystem monitoring and broadcast the aggregated \textit{Health Vector} to the swarm.}
    \label{fig:func_arch}
    \vspace{-4mm}

\end{figure}

\begin{figure*}[htbp]
    \centering
    \includegraphics[width=0.85\textwidth]{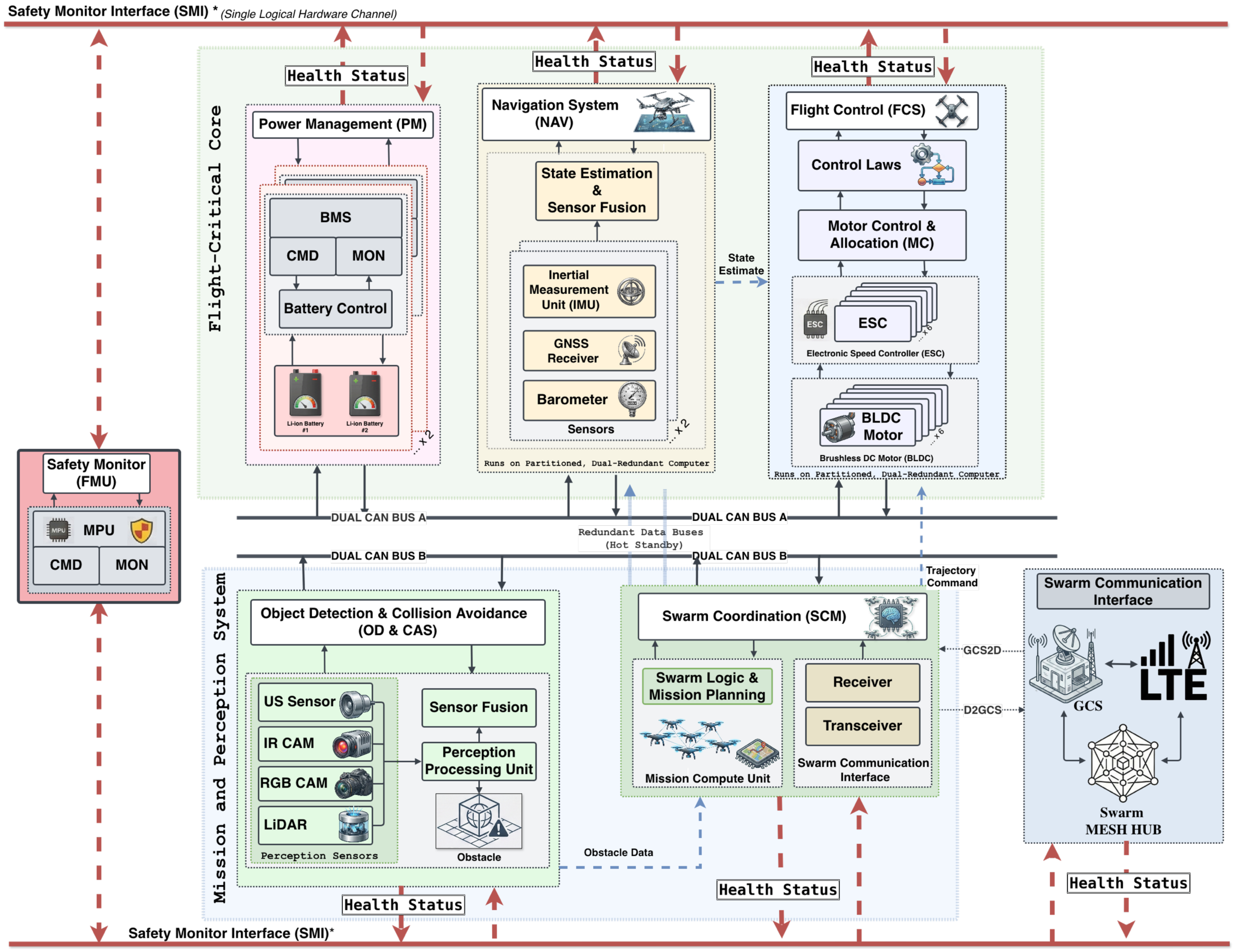}
    \vspace{-3mm}
     \caption{Conceptual block diagram of the proposed fault-tolerant architecture. The system partitions into a \textit{Flight-Critical Core} (DAL C) and a \textit{Mission and Perception System} (DAL D , fail-silent). Redundancy includes dual-channel power and flight computers. The supervisory \textit{Safety Monitor (FMU)} (DAL B) uses a \textit{Command/Monitor (CMD/MON)} architecture to enforce safety contracts via the \textit{Safety Monitor Interface (SMI)}. The SMI is shown at top and bottom for layout clarity; both represent the same logical hardware channel isolating safety traffic from redundant CAN buses.}
     \label{fig:agent_architecture}
     \vspace{-4mm}

\end{figure*}

The foundation of the swarm relies on the integrity of each agent. Assuming a physical platform with actuator redundancy (e.g., a hexacopter), the logical architecture partitions into two subsystems with distinct DALs to maintain fault containment. Two complementary architectural views illustrate this split in Fig.~\ref{fig:func_arch} and Fig.~\ref{fig:agent_architecture}.

As shown in Fig.~\ref{fig:func_arch}, the functional architecture enforces a \textit{Command/Monitor} relationship. This logic is physically mapped in the conceptual architecture shown in Fig.~\ref{fig:agent_architecture}. The first subsystem is the \textit{Flight-Critical Core} (DAL C), which contains all modules essential for safe flight. To achieve fault tolerance, these functions execute on a dual-channel redundant flight computer. The second subsystem is the \textit{Mission and Perception System}, which handles non-deterministic swarm coordination. A single-channel mission computer hosts this system and is treated as a \textit{fail-silent} component. This hierarchical partitioning prevents propagation of software faults in the mission system to the flight core.

The logic modules within these subsystems are:

\begin{itemize}
    \item \textbf{Fault Management Unit (FMU):} The hardware implementation of the RTA \textit{Safety Monitor}. It receives health vectors from all modules and enforces Safety Contracts (Sec.~\ref{ssec:formal_interface}). To protect against OS or middleware failures, the FMU is an independent hardware-isolated unit, extending the System-Level Simplex concept~\cite{Bak2009}. Its logic is verifiable and deterministic. The FMU is assigned DAL B, consistent with the Hazardous classification of F1 in the FHA. The required hardware integrity ($\lambda_{monitor\_hw} \approx 10^{-8}$/hr) is achieved using a CMD/MON architecture, reducing common-mode software faults and improving reliability beyond DAL C.

    \item \textbf{Navigation System (NAV):} Is a DAL C software module executed on the dual-channel flight computers. To satisfy [SSR-FLT-01], it fuses data from redundant IMUs and GNSS receivers to produce a high-integrity state estimate. Its internal Kalman filter provides a final layer of fault detection and exclusion to prevent H-02.
    
    \item \textbf{Flight Control System (FCS):} The core DAL C software module that computes actuator commands. It runs on the flight computer within a partitioned RTOS (implementing ARINC 653 concepts for temporal and spatial isolation). Accepts supervisory commands from the FMU, such as switching to a degraded control allocation matrix.
    
    \item \textbf{Motor Control (MC):} The interface between abstract FCS commands and physical actuators. It translates thrust commands into PWM signals and handles switching to degraded control maps under FCS direction.

    \item \textbf{Power Management (PM):} A DAL C component providing power through a dual-redundant architecture. It includes the Battery Management System (BMS) that feeds telemetry ($V_{batt}$) directly to the FMU for contract evaluation.
    
    \item \textbf{Swarm Coordination Module (SCM):} The primary DAL D component for mission-level tasks. It executes non-deterministic swarm algorithms. The FMU monitors the SCM via a \textit{heartbeat mechanism}. A fault in the SCM (e.g., software crash, missed heartbeat) causes the FMU to declare it failed and command the flight-critical core to a safe contingency state, satisfying \textit{[SSR-FTR-02]}. The SCM receives the aggregated Health Vector from the FMU to broadcast to the swarm mesh.
    
    \item \textbf{Object Detection \& Collision Avoidance (OD-\&CAS):} This DAL D module runs on the mission computer. Outputs are untrusted and bounded by the FMU.
    
    \item \textbf{Communications (COM):} Redundant D2D/D2GCS radios. Powered by PM for availability, logically interfaced with the Mission System. All channels use authenticated encryption to mitigate H-07. The Swarm Mesh Hub is an optional ground relay extending D2D mesh coverage for large-area SAR. It is treated as an untrusted infrastructure component; failures are handled by $\mathcal{C}_{comm}$,  which enforces RTB on mesh loss, whether due to radio faults or hub unavailability.
    
\end{itemize}

\subsubsection{Safety Monitor Communication}
\label{sssec:fmu_comm}

The system safety relies on the isolation of the FMU achieved via the \textbf{Safety Monitor Interface (SMI)}, a dedicated physical channel. This prevents failures on the primary \textit{DUAL CAN BUS}---such as a message flood---from blocking safety commands.

Communication on the SMI is bidirectional:
\begin{itemize}
    \item \textbf{From Modules to FMU:} Critical modules report their health vector ($H_i$), based on internal monitoring (e.g., voltage levels, RPM mismatch, watchdog timeouts, error counters).
    \item \textbf{From FMU to Modules:} The FMU issues high-priority, overriding commands to enforce the Safety Contract.
\end{itemize}

For example, consider a healthy FCS with an actuator fault:
\begin{enumerate}
    \item The FCS detects a persistent kinematic error and reports this raw atomic data to the FMU via the SMI.
    \item The FMU evaluates this data against the Propulsion Contract ($\mathcal{C}_{prop}$), identifies the breach, and transitions the internal state to $\sigma_{prop} \leftarrow \text{DEGRADED}$.
    \item The FMU simultaneously sends an override command (e.g., \texttt{ENFORCE\_SAFE\_ENVELOPE}) back to the FCS and propagates the updated $H_i$ to the SCM.
\end{enumerate}
This creates a fault management loop that remains operational even if the primary data network fails.

% --------------------------------------------------------------------
% SUBSECTION B: SWARM-LEVEL COORDINATION AND RESILIENCE
% --------------------------------------------------------------------
\subsection{Swarm-Level Coordination and Resilience}
\label{ssec:swarm_arch}
While the individual agent architecture handles internal faults, swarm-level resilience manages the failure or loss of entire agents. The swarm coordination logic is directly responsive to the health status provided by each drone's FMU.

\begin{figure*}[!htbp]
    \centering
    \includegraphics[width=0.95\textwidth]{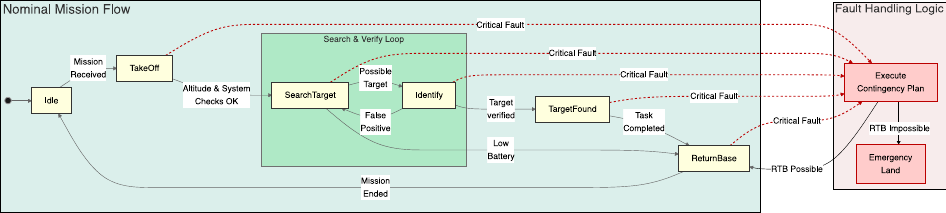}
    \vspace{-3mm}
    \caption{Finite State Machine (FSM) for an individual agent. Any fault in a nominal state triggers transition to Fault Handling Logic.}
    \label{fig:fsm}
    \vspace{-2mm}
\end{figure*}

The behavior of each agent is governed by the Finite State Machine (FSM) shown in Fig.~\ref{fig:fsm}. The FSM design separates the \textit{Nominal Mission Flow} and the \textit{Fault Handling Logic}. Upon a ``Critical Fault'' event triggered by its internal FMU, an agent transitions from any nominal state to the \textit{Execute Contingency Plan} state. Based on the severity of the fault reported by the FMU, one of two paths is chosen:

\begin{itemize}
    \item \textbf{Return to Base (RTB) Possible:} If the FMU indicates the agent retains sufficient navigation and control capability, the FSM transitions to the \textit{Return Base} state for a controlled recovery.
    \item \textbf{RTB Impossible:} If the fault is too severe for a controlled return (e.g., loss of navigation), the FSM transitions to an \textit{Emergency Land} state to minimize risk.
\end{itemize}
The agent broadcasts its degraded status to the swarm to enable a coordinated collective response.

This FSM-driven fault handling enables a two-tiered swarm response that satisfies the SSRs:
\begin{enumerate}
    \item \textbf{Graceful Degradation:} If a drone enters a contingency state but is capable of controlled flight (e.g., the ``RTB Possible'' path), it remains a usable asset. The swarm coordination module re-tasks it to a less critical role, such as a high-altitude communication relay.
    \item \textbf{Agent Loss Response:} If a drone executes an emergency landing or ceases communication entirely (detected via heartbeat timeout), the swarm coordination module removes it from the active roster. To satisfy \textbf{[SSR-MSN-02]}, the module initiates a task re-allocation protocol, redistributing the lost agent's search area among the remaining healthy agents.
\end{enumerate}
A low-battery condition detected during RTB triggers a controlled transition to mission termination. This transition is managed by the $\mathcal{C}_{prop}$ contract via $V_{batt}$ monitoring and does not constitute a hazardous event, provided the agent maintains controlled flight to landing. To mitigate collision risks during a simultaneous swarm-wide return, the \textit{Return Base} state executes a static deconfliction strategy, assigning pre-allocated altitude corridors to each agent based on their ID.

\vspace{-2mm}
% --------------------------------------------------------------------
% SUBSECTION C: SWARM-LEVEL INTERFACE
% --------------------------------------------------------------------
\subsection{Swarm-Level Interface: The Cost Function}
To demonstrate that the architecture is agnostic to specific planning algorithms, we define a standard interface for task allocation based on a cost minimization function. We assume a generic planner (e.g., Market-Based) where agent $i$ bids a cost $J_{ij}$ to perform task $j$. The Swarm Manager (DAL D) receives the Health Vector $H_i$ from the Safety Monitor and incorporates it into the cost function to satisfy \textbf{[SSR-MSN-02]}. We formulate the cost $J_{ij}$ as:
\begin{equation}
J_{ij} = w_d ||p_i - p_{task_j}|| + \mathcal{P}(H_i, \text{TaskType}_j)
\label{eq:cost_function}
\end{equation}
where $p$ denotes position vectors, $w_d$ is a distance weight, and $\mathcal{P}$ is the architectural penalty function enforced by the interface. The penalty function translates the discrete states of the Health Vector into the continuous planning domain:
\vspace{-1mm}
\begin{equation}
\mathcal{P} = 
\begin{cases} 
0 & \text{if } H_i = \text{NOMINAL} \\
\infty & \text{if } H_i = \text{DEGRADED} \land \text{Task}_j \in \text{Agility} \\
\lambda_{deg} & \text{if } H_i = \text{DEGRADED} \land \text{Task}_j \in \text{Relay}
\end{cases}
\end{equation}
where $\lambda_{deg} > 0$ is a finite mission penalty assigned to degraded agents performing relay tasks, reflecting their reduced operational capability. The specific value of $\lambda_{deg}$ is a tuning parameter of the swarm planner. It is set such that a degraded agent remains competitive for relay tasks but is systematically deprioritized relative to healthy agents ($\lambda_{deg} \ll \infty$). Its calibration is a planner-level design decision outside the scope of this architectural framework.

This mathematical interface results in a ``Degraded'' agent being priced out of high-agility tasks (infinite cost) while remaining competitive for static relay tasks (finite penalty $\lambda_{deg}$). This realizes fail-operational behavior without complex rule-based logic. The cost function operates within the DAL~D Swarm Coordination Module and is therefore bounded by the DAL B Safety Monitor. A failure of the cost function to correctly penalize degraded agents is treated as an F4-class failure (Table~\ref{tab:fha}) and is contained by the FMU's trajectory validation prior to execution.

\section{Architectural Analysis}
\label{sec:evaluation}
% ====================================================================
% SECTION V: ARCHITECTURAL ANALYSIS AND THEORETICAL COMPLIANCE ARGUMENT
% ====================================================================

Given the conceptual design phase of this work, the framework is assessed through three complementary systems engineering methods: (1) \textbf{Architectural Traceability Analysis}, to verify that each identified hazard is mitigated by a specific architectural control; (2) \textbf{Markov Reliability Modeling}, to establish the theoretical compliance boundary and derive the Safety Monitor coverage requirement; and (3) a \textbf{ConOps Scenario Walkthrough}, to verify the logical consistency and timing of the fault response chain.

% --------------------------------------------------------------------
% SUBSECTION A: TRACEABILITY & QUALITATIVE FTA
% --------------------------------------------------------------------
\subsection{Traceability and Qualitative Analysis}
\label{ssec:traceability}

A primary requirement of ARP4754B is the traceability of hazards to architectural mitigations. Table~\ref{tab:mitigation} demonstrates this link. Every hazard identified in the FHA (Section~\ref{sec:methodology}) is mapped to a specific hardware or software control.

\begin{table}[htbp]
\caption{Traceability of Hazards to Architectural Mitigations}
\vspace{-5mm}
\label{tab:mitigation}
    \begin{center}
    \resizebox{\columnwidth}{!}{%
        \begin{tabular}{|p{2.5cm}|p{8.5cm}|}
        \hline
        \textbf{Hazard ID} & \textbf{Mitigating Architectural Features (from Section IV)} \\
        \hline
        \hline
        \textbf{H-01} Collision & \textbf{Layered Defense:} Swarm planner (DAL D) provides strategic deconfliction; Reactive OD\&CAS (DAL D) provides tactical avoidance; Flight Core (DAL C) enforces geofencing limits. \\
        \hline
        \textbf{H-02} Crash & \textbf{Hardware Isolation:} Dual-redundant Flight Core (DAL C); Independent Safety Monitor (FMU, DAL B) prevents SCM from commanding unsafe states via the Safety Contract. \\
        \hline
        \textbf{H-03} Mission Fail & \textbf{Swarm Reconfiguration:} Task re-allocation logic (SSR-MSN-02) triggered by Health Vector $H_i(t)$ updates via SMI. \\
        \hline
        \textbf{H-06} Link Loss & \textbf{Diversity:} Redundant D2D/D2GCS radios. Loss of GCS triggers mesh-based autonomous replanning; Loss of Mesh (D2D) triggers independent 'Return-to-Base'. \\
        \hline
        \textbf{H-07} Unauth. Access & \textbf{Encryption:} Authenticated encryption on all SMI and COM links (SSR-SEC-01). \\
        \hline
        \end{tabular}
    }
    \end{center}
    \vspace{-5mm}
\end{table}

\begin{figure}[ht]
    \centering
    \includegraphics[width=\columnwidth]{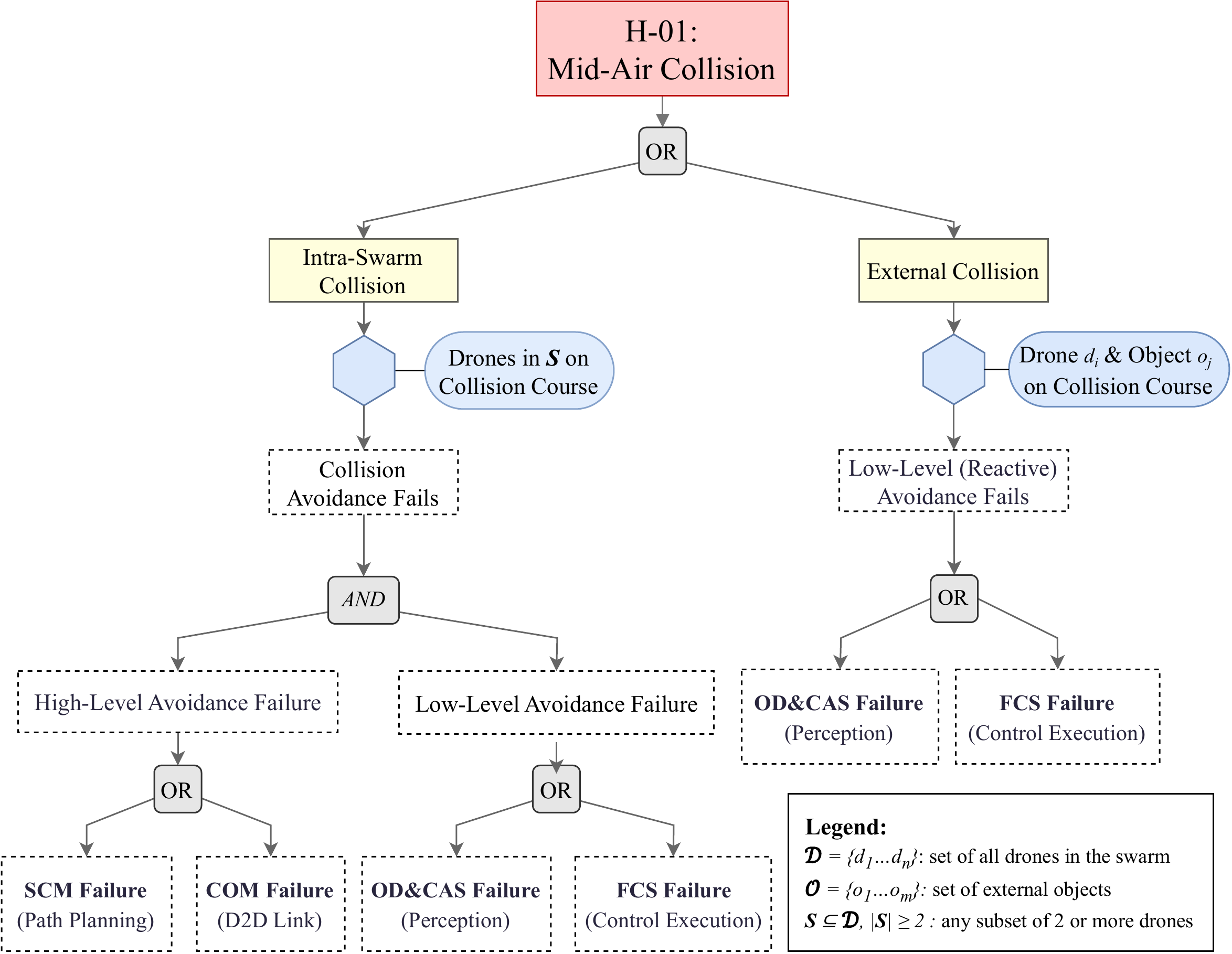}
    \vspace{-5mm}
    \caption{Qualitative Fault Tree Analysis (FTA) for the 'Mid-Air Collision' (H-01) swarm-emergent hazard. The tree demonstrates the system's multi-layered safety architecture for intra-swarm collision, requiring failures in both high-level (cooperative) and low-level (reactive) avoidance layers.}
    \label{fig:fta_h01}
    \vspace{-2mm}

\end{figure}

To verify defenses against swarm-emergent hazards, Fig.~\ref{fig:fta_h01} shows the Qualitative Fault Tree Analysis (FTA) for a mid-air collision (H-01). The logic demonstrates the architecture's functional redundancy. The top-level OR gate partitions the hazard into 'Intra-Swarm Collision' and 'External Collision'. For an intra-swarm collision to occur, an AND gate dictates that drones must be on a collision course concurrently with a 'Collision Avoidance Fails' event. Furthermore, an AND gate confirms that both the \textit{High-Level Avoidance} (Cooperative Swarm Planning) and \textit{Low-Level Avoidance} (Reactive Sensors/FCS) must fail simultaneously.

% --------------------------------------------------------------------
% SUBSECTION B: MARKOV ANALYSIS
% --------------------------------------------------------------------
\subsection{Quantitative Reliability Analysis}
\label{ssec:quant_analysis}

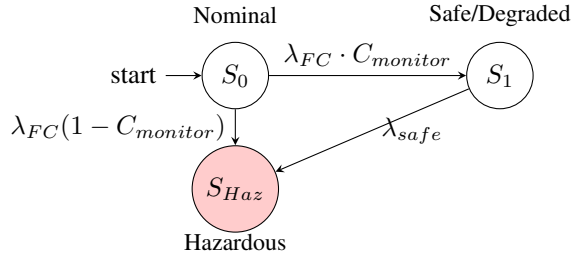
\begin{figure}[ht]
    \centering
    \begin{tikzpicture}[>=stealth, node distance=2.5cm, auto]
        % Nodes
        \node[state, initial] (S0) {$S_0$};
        \node[state, right of=S0, node distance=3.5cm] (S1) {$S_1$};
        \node[state, below of=S0, node distance=1.5cm, fill=red!20] (S2) {$S_{Haz}$};
        
        % Transitions
        \path[->] 
            (S0) edge node {$\lambda_{FC} \cdot C_{monitor}$} (S1)
            (S0) edge node[left] {$\lambda_{FC} (1-C_{monitor})$} (S2)
            (S1) edge node[right] {$\lambda_{safe}$} (S2);
            
        % Labels
        \node[above of=S0, node distance=0.8cm] {\small Nominal};
        \node[above of=S1, node distance=0.8cm] {\small Safe/Degraded};
        \node[below of=S2, node distance=0.7cm] {\small Hazardous};
    \end{tikzpicture}
    \vspace{-2mm}
    \caption{Markov Chain model of the architecture. $S_0$: Nominal operation. $S_1$: Successfully contained fault (Fail-Operational). $S_{Haz}$: Uncontained failure. System safety depends on Monitor Coverage ($C_{monitor}$). $\lambda_{safe}$ is the secondary failure rate from degraded state $S_1$. During short RTB recovery, this transition is dominated by $\lambda_{FC}$ and bounded as described in the text.}
    \label{fig:markov}
    \vspace{-1mm}
\end{figure}

To demonstrate compliance with the Hazardous safety target ($< 10^{-7}$ per flight hour for H-02), we model the system dynamics using a Continuous Time Markov Chain (CTMC). Unlike static Fault Trees, this model explicitly captures the architecture's dependence on the Safety Monitor's diagnostic coverage ($C_{monitor}$). The state space, in Fig.~\ref{fig:markov}, consists of:
\begin{itemize}
    \item \textbf{State 0 (Nominal):} All systems functional. The Swarm Coordinator (DAL D) manages the mission.
    \item \textbf{State 1 (Degraded/Safe):} A hardware fault occurred, but the Safety Monitor (FMU) detected it and enforced a contingency (e.g., Hex-to-Quad control allocation or RTB). This represents a fail-operational state, as the system maintains controlled flight and swarm coordination.
    \item \textbf{State 2 (Hazardous):} Uncontrolled crash. This occurs if the Monitor fails to cover the fault (Lack of Coverage) or if the hardware fails  without redundancy.
\end{itemize}

The probability of a Hazardous state is dominated by the Coverage Factor ($C_{monitor}$) of the Safety Monitor. The flight computer operates in a Command/Monitor (CMD/MON) configuration, so the CMD channel generates actuator commands while the MON channel verifies their correctness in parallel. An undetected hazardous failure requires the CMD to produce an error that propagates to the actuators, the rate of which is captured by $\lambda_{FC}$, rather than either channel independently failing. Using a first-order approximation for rare events, the steady-state hazard rate $\lambda_{haz}$ is formulated as:
\begin{equation}
    \lambda_{haz} \approx \lambda_{FC} \cdot (1 - C_{monitor}) + \lambda_{monitor\_hw}
    \label{eq:hazard_rate}
\end{equation}
Where:
\begin{itemize}

    \item $\lambda_{FC} \approx 10^{-4}$/hr is adopted as a conservative parametric estimate for a single-channel flight computer. This value is consistent with failure rate ranges for aerospace-grade embedded processors~\cite{mil_hdbk_217} and is used here as a bounding assumption. The sensitivity of the compliance result to this parameter is analyzed in Table~\ref{tab:sensitivity}.
    
    \item $\lambda_{monitor\_hw} \approx 10^{-8}$/hr represents the undetected hardware failure rate for the FMU under its DAL~B assignment. A hazardous FMU failure requires both a hardware fault in the FMU \textit{and} a failure of the internal CMD/MON monitor to detect it. This value corresponds to the integrity target achievable with a CMD/MON processor architecture~\cite{ti_tms570}, and is treated here as a design requirement rather than a measured value.
    
\end{itemize}

The transition from the degraded state $S_1$ to the hazardous state $S_{Haz}$ via secondary failure ($\lambda_{safe}$) is omitted from the bounding approximation, justified as follows: upon entering $S_1$, the FMU enforces the Safety Contract, restricting the agent to a reduced velocity envelope and initiating a Return-to-Base maneuver. The exposure time in $S_1$ is bounded by the RTB maneuver duration $t_{RTB}$, which for SAR platforms operating at SAIL~IV distances is on the order of minutes. The probability of a secondary hazardous failure during this window is bounded by $\lambda_{FC} \cdot t_{RTB}$. For $\lambda_{FC} = 10^{-4}$/hr and $t_{RTB} \le 10$ min $= 1/6$ hr, this yields $P_{secondary} \le 1.7 \times 10^{-5}$, which is below the Hazardous threshold and confirms the omission does not affect the bounding result.

Communication failure is architecturally bounded by the $\mathcal{C}_{comm}$ contract (Section~\ref{ssec:formal_interface}), which enforces a deterministic Return-to-Base maneuver upon link degradation, preventing communication loss from directly inducing an uncontrolled crash (H-02). For the overall hazard rate to remain dominated by $C_{monitor}$ rather than $\lambda_{comm}$, the redundant communication subsystem must satisfy the design requirement $\lambda_{comm} \ll \lambda_{FC} = 10^{-4}$/hr. This is a system-level specification to be verified during hardware selection and is consistent with the reliability targets achievable by redundant mesh radio architectures in the UAV domain.

Rather than assuming a coverage value, we solve for the minimum required coverage to satisfy the Hazardous failure-condition target ($P_{fail} < 10^{-7}$). Solving Eq.~\ref{eq:hazard_rate} for $C_{monitor}$:
\begin{equation}
\resizebox{0.86\columnwidth}{!}{$
    10^{-4} \cdot (1 - C_{monitor}) + 10^{-8} < 10^{-7} \Rightarrow C_{monitor} > 0.9991
$}
\end{equation}

\vspace{-4mm}
\begin{table}[htbp]
\caption{Sensitivity of Required Safety Monitor Coverage to Flight Computer Failure Rate $\lambda_{FC}$}
\vspace{-3mm}
\label{tab:sensitivity}
    \begin{center}
    \resizebox{0.9\columnwidth}{!}{%
        \begin{tabular}{|c|c|p{4.1cm}|}
        \hline
        \textbf{$\lambda_{FC}$ (per flight hour)} & \textbf{Required $C_{monitor}$} & \textbf{Achievable Implementation} \\
        \hline
        \hline
        $10^{-3}$ & $> 0.99991$ & DAL A CMD/MON architecture \\
        \hline
        $10^{-4}$ & $> 0.9991$ & DAL B CMD/MON architecture \\
        \hline
        $10^{-5}$ & $> 0.991$ & DAL C architecture (e.g. lockstep strategy for detecting random faults) \\
        \hline
        $10^{-6}$ & $> 0.91$ & Software monitor sufficient \\
        \hline
        \end{tabular}
    }
    \end{center}
    \vspace{-2mm}
\end{table}

Table~\ref{tab:sensitivity} demonstrates that the architecture remains compliant across a range of flight computer reliability assumptions. For the baseline assumption of $\lambda_{FC} = 10^{-4}$/hr, a DAL~B CMD/MON implementation of the FMU satisfies the coverage requirement. If higher-integrity flight computers are used ($\lambda_{FC} \le 10^{-5}$/hr), the coverage requirement relaxes, permitting less costly monitoring implementations. By enforcing this requirement through strict hardware isolation, the architecture derives its safety assurance from verifiable physical properties rather than probabilistic software behavior.

% --------------------------------------------------------------------
% SUBSECTION C: COMPARATIVE ANALYSIS
% --------------------------------------------------------------------
\subsection{Comparative Analysis}
\label{ssec:comp_analysis}
Table~\ref{tab:comparison} compares our framework against both traditional fault-tolerant control and modern RTA standards.

\begin{table}[!htpb]
\caption{Comparative Analysis of Architectures}
\label{tab:comparison}
\centering
\resizebox{\columnwidth}{!}{%
    \begin{tabular}{|c|p{3cm}|p{3cm}|p{4cm}|}
        \hline
        \textbf{Criterion} & \textbf{Single-Agent FTC \cite{Nguyen2019}} & \textbf{Std. RTA (ASTM F3269)} & \textbf{Proposed Architecture} \\
        \hline \hline
        \textbf{Scope} & Single Drone Only & Single Drone Only & \textbf{Swarm \& Agent} \\
        \hline
        \textbf{Safety Mechanism} & Algorithmic Redundancy (Kalman Filters) & Simplex Monitor (Switch to Safe Controller) & \textbf{Hardware Monitor + Swarm Reconfiguration} \\
        \hline
        \textbf{Response to Fault} & Adjust Control Laws (Local) & Terminate / Loiter (Local) & \textbf{Propagate Health Vector $H_i(t) \rightarrow$ Global Replan} \\
        \hline
        \textbf{Mixed-Criticality} & No (All software is critical) & Yes (Monitor vs. Complex) & \textbf{Yes (DAL B Monitor / DAL C Flight Core / DAL D Swarm)} \\
        \hline
        \textbf{SWaP \& Scope} & Low (any platform) & Low-Medium (any platform) & \textbf{High - targets $>10$ kg MTOW; excludes micro-UAVs} \\
        \hline
    \end{tabular}%
}
\vspace{-3mm}
\end{table}

The comparison shows that while ASTM F3269 provides the safety monitor concept, it lacks the mechanism to propagate the cause of the switch to the swarm. Our architecture fills this gap by turning the RTA monitor into a source of health data for the collective. This architectural cost is a design trade-off: the SWaP (Size, Weight, and Power) overhead of hardware isolation and redundancy is the mechanism by which DAL~B integrity is achieved without software-only assurance.

% --------------------------------------------------------------------
% SUBSECTION D: CONOPS VERIFICATION
% --------------------------------------------------------------------
\subsection{ConOps Verification: Actuator Degradation}
\label{ssec:usecase}

\begin{figure}[ht]
    \centering
    \includegraphics[width=0.9\columnwidth]{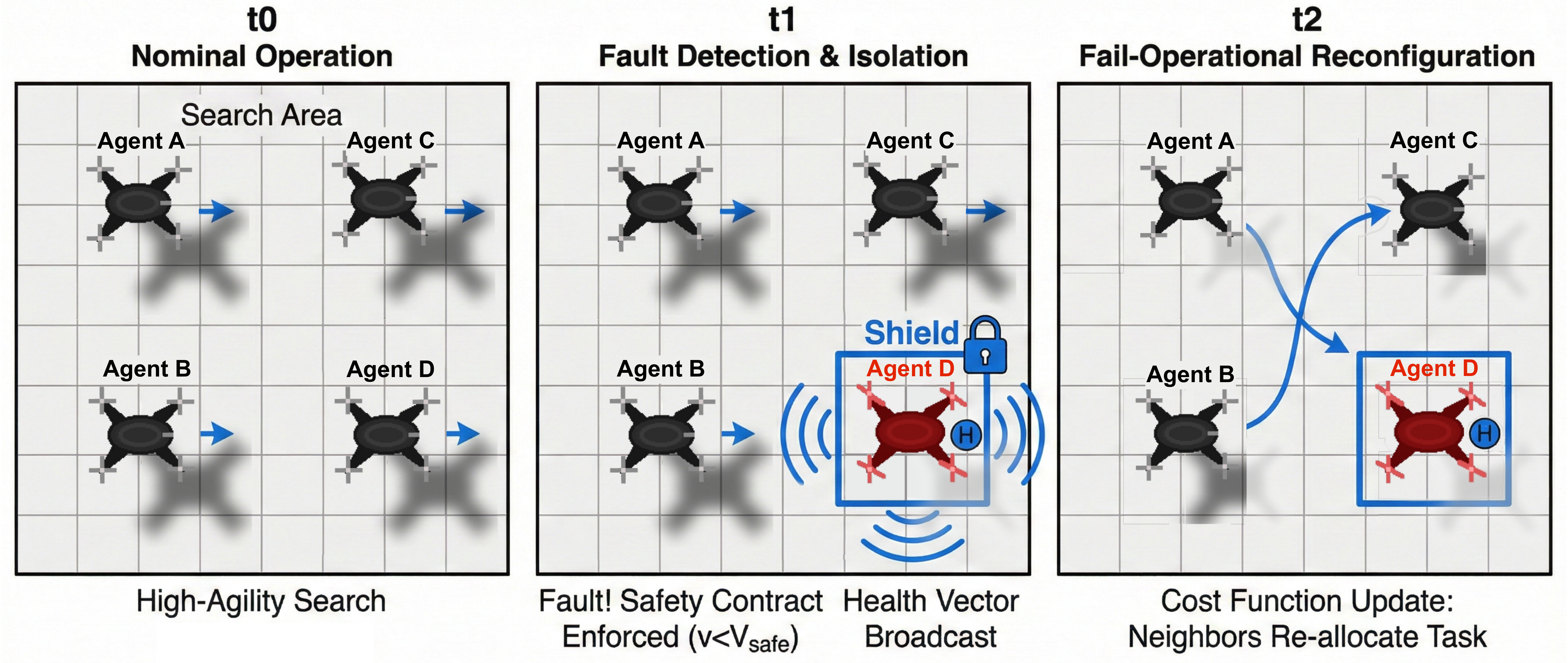} 
    \vspace{-3mm}
    \caption{\textbf{Operational Sequence of the Verifiable Health-Aware Framework.} 
    (\textbf{t0}) Nominal operation: Swarm executes search under AI control. 
    (\textbf{t1}) Fault Detection: Upon hardware fault detection (e.g., motor loss), the Safety Monitor isolates the Agent~$D$ into a safety envelope ($v \le v_{safe}$) and broadcasts a degraded Health Vector $H_D(t)$. 'Shield' denotes the enforcement of the Safety Contract by the FMU.
    (\textbf{t2}) Reconfiguration: The swarm updates the distributed Cost Function (Eq.~\ref{eq:cost_function}), penalizing Agent~$D$ and re-allocating its search sectors to healthy neighbors (e.g., Agents~$A$ and $C$).}
    \label{fig:conops_seq}
    \vspace{-1mm}
\end{figure}

Figure~\ref{fig:conops_seq} illustrates the sequence during a runtime fault. The specific timing and signal propagation of a partial motor failure event are traced to verify compliance with the SSRs.
% ====================================================================
% FIGURE B: OPERATIONAL SEQUENCE (ConOps)
% ====================================================================
\vspace{-2mm}

\begin{enumerate}
    \item \textbf{Fault Onset ($t=0$):} Motor 5 on Agent $D$ suffers a winding short, reducing thrust by 40\%.
    
    \item \textbf{Detection ($t \le 50$\,ms):} Based on the vehicle's time-to-divergence (see $\tau_{max}$ derivation in Section~\ref{ssec:formal_interface}), the FMU correlates the yaw rate divergence (FCS) and current spike (BMS) within 50\,ms.
    
    \item \textbf{Isolation ($t \le 100$\,ms):} The FMU enforces the Propulsion Contract ($\mathcal{C}_{prop}$). It commands the FCS to switch to the "Hex-to-Quad" degraded control allocation matrix and enforces the velocity restriction ($v \le v_{safe}$). It updates the Agent Health Vector: \\
    $H_D(t) = \langle \text{NOMINAL}, \text{DEGRADED}, \text{NOMINAL} \rangle$. This sequence satisfies the timing constraint of Eq.~\ref{eq:latency}: $\tau_{detect} + \tau_{enforce} = 50 + 50 = 100$\,ms $\le \tau_{max}$.
    
    \item \textbf{Propagation ($t \le 200$\,ms):} The SCM reads the updated Health Vector $H_D(t)$ via the Safety Monitor Interface (SMI) and broadcasts it to the swarm mesh network.
    
    \item \textbf{Transient Kinematic Safety ($200 \le t < 500$\,ms):} During replanning, inter-UAS spacing is maintained by the Reactive OD\&CAS layer of trailing agents, operating independently of the global swarm planner. Satisfies the design requirement: the detection-to-avoidance latency $\tau_{OD}$ must be less than the time-to-collision $t_{TTC}$ at minimum operational separation, i.e., $\tau_{OD} \ll d_{min} / v_{rel\_max}$, where $d_{min}$ is the minimum inter-agent separation distance and $v_{rel\_max}$ is the maximum relative approach velocity at the onset of fault. This requirement is a platform-level specification that must be verified during integration and flight testing.
    
    \item \textbf{Decentralized Reallocation ($t \le 500$\,ms):} The Swarm Coordination Modules on neighbor drones (Agents $A, B, C$) receive the vector $H_D(t)$. Executing a decentralized consensus algorithm, they identify that Agent $D$ can no longer hold its precise search trajectory in high wind due to its degraded propulsion state.
    
    \item \textbf{Mission Adaptation ($t > 500$\,ms):} Healthy neighbors adjust their flight paths (indicated by blue arrows in Fig.~\ref{fig:conops_seq}) to cover the operational gap. Agent $D$ transitions to a static relay node role.
\end{enumerate}

Altitude deconfliction during simultaneous RTB is handled by the pre-allocated corridor strategy described in Section~\ref{ssec:swarm_arch}. This trace illustrates how the \textit{Safety Monitor Interface} couples hardware faults (millisecond-scale) with mission adaptation (second-scale), supporting SSR-MSN-02 while maintaining DAL C functional isolation.

\section{Discussion}
\label{sec:discussion}
% ====================================================================
% SECTION VI: DISCUSSION AND LIMITATIONS
% ====================================================================
\vspace{-0.5mm}

The presented approach maps hazards to architectural mitigations to establish a verifiable basis for design assurance. The hierarchical fault management framework enables the swarm to reallocate tasks to degraded resources rather than treating a partially failed drone as a complete loss. This reallocation executes graceful degradation during autonomous operations.

The architecture partitions subsystems by criticality. The DAL~B Safety Monitor provides the highest integrity layer to enforce hardware-isolated containment of DAL~D failures. Flight-critical modules (PM, NAV, FCS) at DAL~C utilize hardware redundancy, whereas the mission-critical SCM operates within a monitored-simplex design at DAL~D. The integration of a physically isolated FMU, dual-redundant flight computers, and redundant data buses introduces a measurable payload penalty. Consequently, this architecture targets payload-capable UAVs (e.g., platforms $>10$~kg MTOW) operating in SAIL~IV environments. Strict SWaP constraints preclude hardware replication in micro-UAV swarms. In the event of an SCM failure, the FMU enforces safe state transitions independent of mission objectives.

\vspace{-0.5mm}
\subsection{Design Assumptions and Limitations}
This work acknowledges limitations based on its current Technology Readiness Level (TRL). First, the quantitative reliability results (Section~\ref{ssec:quant_analysis}) are model-based. The evaluation that the Hazardous failure-condition target is theoretically
achievable for our SAIL~IV scenario relies on the assumption that the Safety Monitor achieves a diagnostic coverage of $C_{monitor} > 0.9991$. While this metric aligns with DAL~A/B CMD/MON hardware implementations, verifying this coverage in practice requires targeted fault injection campaigns outside the scope of this architectural proposal.

Second, the model assumes independence between the Safety Monitor and the Mission Computer. In a physical implementation, shared resources (such as power buses or clock lines) introduce the potential for Common Cause Failures (CCF). To transition this theoretical framework into a practical Proof-of-Concept (PoC), future work targets the Hardware-in-the-Loop (HIL) verification of the FMU using representative FPGA hardware to verify the timing guarantees of the Safety Monitor Interface (SMI) under saturation conditions.

Finally, while this framework defines the formal interface (the Cost Function) between agent health and swarm logic, the effectiveness of the reconfiguration depends on the convergence speed of the distributed planner. The verification and validation (V\&V) of non-deterministic replanning algorithms remains an open research challenge~\cite{Forsberg2020}.
\vspace{-1mm}

\section{Conclusion}
\label{sec:conclusion}
% ====================================================================
% SECTION VII: CONCLUSION
% ====================================================================

This paper introduced a verifiable health-aware architectural framework for autonomous swarms. Applying the ARP4754/ARP4761 framework derived a mixed-criticality architecture where a hardware-isolated Safety Monitor (DAL~B) enforces formal safety contracts on non-deterministic swarm logic (DAL~D). Integrating Run-Time Assurance (RTA) with swarm reconfiguration links flight-critical component reliability (DAL~C) to multi-agent resilience. Markov analysis establishes that the Hazardous failure-condition target ($P_{fail} < 10^{-7}$/hr) is theoretically achievable for our SAIL~IV scenario if the Safety Monitor achieves $C_{monitor} > 0.9991$. This requirement is consistent with DAL~B hardware architectures, such as CMD/MON architectures ($\sim 10^{-7}$/hr). By defining boundaries between flight-critical systems and complex swarm algorithms, the framework supports the certification of autonomous operations. Future work targets Hardware-in-the-Loop verification of the Safety Monitor using representative FPGA hardware. This verification focuses on the timing guarantees of the Safety Monitor Interface (SMI) and the fault injection response relative to Safety Contract bounds. %These steps are required to establish a certifiable RTA boundary for search and rescue operations.
\vspace{-1mm}

\section*{acknowledgment}
This research was funded by the Horizon Europe research and innovation program of the European Union and the Chips Joint Venture under GA No. 101194287, NexTArc (Next Generation Open Innovations in Trustworthy Embedded AI Architectures for Smart Cities, Mobility, and Logistics).

\bibliographystyle{IEEEtran}
\bibliography{references}

\end{document}